\documentclass[twocolumn,showpacs,preprintnumbers,amsmath,amssymb,prl]{revtex4}

\usepackage{graphicx}
\usepackage{bm}

\newcommand{\EQA}[1] {\begin{eqnarray}#1\end{eqnarray}}
\newcommand{\PD} {\partial}
\newcommand{\EQAMK} {&\!\!\!}

\newcommand{\Ps} {P_s}

\begin{document}

\title{A new model of turbulent relative dispersion: a
self-similar telegraph equation
based on persistently separating motions
}

\author{Takeshi OGASAWARA}
\email{ogasawara@kyoryu.scphys.kyoto-u.ac.jp}
\author{Sadayoshi TOH}
\email{toh@scphys.kyoto-u.ac.jp}
\affiliation{
Division of Physics and Astronomy, Graduate School of Science, Kyoto
University, Kyoto 606-8502, Japan}

\date{\today}

\begin{abstract}
 Turbulent relative dispersion is studied theoretically
 with a focus on  the evolution of probability distribution of
 the relative separation of 
 two passive particles.
 A finite separation speed and a finite correlation of relative velocity,
 which are crucial for real turbulence, are implemented to a master
 equation by multiple-scale consideration.
 A telegraph equation with 
 scale-dependent coefficients is derived in the continuous limit.
 Unlike the conventional case, the telegraph equation has a similarity
 solution bounded by the maximum separation.
 The evolution is characterized by two parameters:
 the strength of persistency of separating motions and
 the coefficient of the drift term.
 These parameters are connected to Richardson's
 constant and, thus, expected to be universal. The relationship between
 the drift term
 and coherent structures is discussed for 
 two 2-D turbulences.
\end{abstract}

\maketitle

Turbulent transport and mixing underlie 
wide range of phenomena from star formations \cite{MK2004} to coffee in
a cup.
However, the mechanisms of their significant deviation from
molecular counterparts
are not well understood yet even in such a simple case as
relative dispersion of passive particles.

In the inertial subrange, reflecting the scaling 
law of turbulent  velocity fluctuation,
well-known Richardson's $t^3$ law is realized 
\cite{R1926,MY1975}.
For the probability density
function (PDF)  of separation $r$, $P(r,t)$,
quite a few
models and theories
consistent with Richardson's law
have been proposed \cite{S2001}.
Besides recent progresses of particle tracking techniques and numerical
simulations enable
direct experimental investigations of the separation PDF
\cite{EXPS,BS2002a,GV2004}.
Although accuracies of the experiments are not enough,
most of their results are so far close to
the prediction of Richardson's diffusion equation \cite{R1926}:
\EQA{
\frac{\PD P}{\PD t}
=
\frac{\PD}{\PD r}
\left[
K(r) r^{d-1}
\frac{\PD}{\PD r}
\left(
\frac{P}{r^{d-1}}
\right)
\right],
\label{eq:Richardson}
}
where $K(r)\propto r^{4/3}$ in the Kolmogorov scaling \cite{MY1975}
and $d$ is the spatial dimension.
This closeness indicates Eq.\ (\ref{eq:Richardson}) is a good basis for
description of turbulent relative dispersion.
Eq.\ (\ref{eq:Richardson}) ought to be
exact only if the relative velocity 
is $\delta$-correlated in time \cite{S2001,S1999}.
However,
there are spatio-temporal correlations in turbulent flows
as implied by existence of coherent structures.
This indicates relative velocity cannot be $\delta$-correlated in time.
To resolve the inconsistency,
Eq.\ (\ref{eq:Richardson}) has to be extended to include these
correlations.
As mentioned in detail below,
we treat these correlations as those not in time but in scale-space.
That is, we focus on correlations between scales $r$ and $\rho r$,
where $\rho$ is a scale multiplier.
This treatment is appropriate for describing self-similarity.
Employing multiple-scale consideration with correlation in scale,
we derive a telegraph equation
with scale-dependent coefficients, Eq.\ (\ref{eq:T-Model}),
and obtain a similarity solution,
which never coincides with
Richardson's one markedly in the tail part
even in the long time limit.

\begin{figure}[b]
 \includegraphics[width=7.2cm,trim=0 0 0 2.0cm,clip]{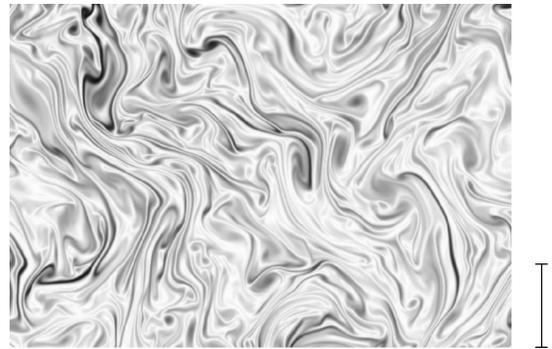}
 \caption{\label{fig:vorticity-snapshot}
 A snapshot of vorticity field of 2D-FC turbulence.
 Shading represents intensity of vorticity.
 The scale denoted by the straight line is $100\eta_\theta$
 and in the inertial subrange, where
 $\eta_\theta$ is the thermal Kolmogorov scale \cite{TM2003}.
 }
\end{figure}
Coherent structures observed in turbulence must share their origin 
with  finite correlation and self-similarity,
so that particles appear to be advected according to coherent structures.
For example, in two-dimensional (2D) inverse cascade
(IC) turbulence,
particles separate step-by-step through nested cat's eye vortices
with scattering by stagnation points \cite{DV2003,GV2004};
in 2D free convection
(FC) turbulence,
particles separate through advection by stretching and folding plumes
(Fig.\ \ref{fig:vorticity-snapshot}) \cite{TM2003}.
Clearly, separation processes in these two systems are different.
However, despite these differences,
effects of coherent structures
on relative dispersion appear in the same way, i.e.,
persistent separation:
persistent expansion and compression of
a relative separation.
Sokolov {\it et al.}\ model such motions based on L\'evy walk.
Their model consists of persistent separation ceased
by probabilistic turn in direction
\cite{S1999,SKB2000}.
They
introduced also the persistent parameter $\Ps$,  the ratio of 
the correlation length 
to the scale.

The telegraph model was introduced to implement 
(I) a finite diffusion speed,  and (II) a finite correlation time
to the diffusion process \cite{G1951,B2003}. It has been widely applied 
to various diffusion phenomena
from molecular diffusion \cite{B2003}
to population dynamics \cite{H1993}.

To satisfy the self-similarity of turbulence,  (I) and (II)
are extended to be scale-dependent according to the following scaling
assumptions: $v(r) = Ar^{1-g}$ and  $T_c(r) = r/v(r)=A^{-1}r^g$, 
where $v(r)$ and  $T_c(r)$  are relative velocity and characteristic time,
 $A$ a constant \footnote{In the case of the Kolmogorov scaling,
$A=C_K\varepsilon^{1/3}$, where $C_K$ and $\varepsilon$
are the Kolmogorov constant and the energy dissipation rate,
respectively.
}, and $g$ a scaling exponent \footnote{$g=2/3$ and $2/5$ for 2D-IC and -FC
turbulences \cite{TS1994}, respectively.}.

First, we  focus on a scale $\hat{r}$ in the inertial subrange.
The correlation length and time  for a separation
to be expanded or
compressed persistently are defined as 
$\Ps^{\pm}\hat{r}$ and $\Ps^{\pm}T_c(\hat{r})$, respectively. 
We assume $\Ps^{\pm}$ is
the order of unity, and then, $\rho_\mathrm{out}=1 + \Ps^{\pm}$. We call
this scale outer.

We consider the evolution of probability density of the separation in a
small region around $\hat{r}$
where its extent is much smaller than $\Ps^{\pm}\hat{r}$. Since  
$\Ps^{\pm}\hat{r}$ and $\Ps^{\pm}T_c(\hat{r})$ are regarded as
constants, we can apply the
approach deriving the telegraph equation to this small region \cite{G1951}.
We call this scale inner.

We divide the inner region  into
shells defined as $[r_n, r_{n+1})$ where $r_n=\hat{r}\xi^n$, $n$ is
integer, and $\xi$ is
close to unity, that is, $\rho_\mathrm{in}=\xi^n $.
Since $\rho_\mathrm{in}-1 \ll \rho_\mathrm{out}-1$,
we deal with two scales in scale-space.
To take into a finite speed, pass-through time $\tau_n$ of the $n$-th shell,
the time for a relative separation to expand (compress) through the
$n$-th shell, is defined as follows:
\EQA{
\tau_n = \frac{\Delta r_n}{v(r_n)}
= \frac{r_{n+1}-r_n}{v(r_n)}
\approx \frac{\gamma r_n}{v(r_n)},
\label{eq:def-tau}
}
where $\gamma=\log\xi$ ($\gamma\ll1$),
and $v(r_n)$ is the relative velocity at a spatial
scale $r_n$.
This relation means $O(\tau_n)=O(\gamma)$, which is the key difference
from the diffusion equation case including Richardson's equation.

We introduce probabilities $Q_{n}^{\pm}$:
the probability for a relative separation to be 
expanding ($Q_{n}^{+}$) or compressing ($Q_{n}^{-}$) in the $n$-th shell.
Transition probabilities $\Delta p^\pm(r_n)$,
the probability from expansion to compression ($+$) and the opposite
($-$)
during $\tau_n$, must be self-similar
in outer scale.
Therefore the simplest form of $\Delta p^\pm(r_n)$ is
\EQA{
\Delta p^{\pm}(r_n) = \frac{\tau_n}{\Ps^\pm T_c(\hat{r})}
=
\frac{\lambda^{\pm}}{T_c(\hat{r})}\tau_n,
\label{eq:def-transition-prob}
}
where $\Ps^\pm T_c(\hat{r})$ are correlation times depending on the direction and 
 $\lambda^\pm=1/\Ps^\pm$.
Note that
$\Ps^{\pm}T_c(\hat{r})$ are outer scale
and, thus, constants in the inner region.
This form was first given by Sokolov {\it et al.}\ \cite{SKB2000}.
Unlike their model,
we assign different values to $\lambda^\pm$ 
to represent  difference of persistency between expansion and compression.

Based on above
considerations,
we construct
master equations for $Q_n^+(t)$ and $Q_n^-(t)$ as follows:
 \EQA{
 Q_n^+(t+\tau_{n-1})
 =
 \left(1-\frac{\lambda^+}{T_c(\hat{r})}\tau_{n-1}\right) Q_{n-1}^+(t)
 \qquad\quad
 \nonumber
 \\
 \qquad\quad
 + \frac{\lambda^-}{T_c(\hat{r})}\tau_{n-1} Q_{n-1}^-(t)
 + \left(1-\frac{\tau_{n-1}}{\tau_n}\right) Q_n^+(t),
 \label{eq:Q+}
 }
 \EQA{
 Q_n^-(t+\tau_{n+1})
 =
 \left(1-\frac{\lambda^-}{T_c(\hat{r})}\tau_{n+1}\right) Q_{n+1}^-(t)
 \qquad\quad
 \nonumber
 \\
 \qquad\quad
 + \frac{\lambda^+}{T_c(\hat{r})}\tau_{n+1} Q_{n+1}^+(t)
 + \left(1-\frac{\tau_{n+1}}{\tau_n}\right) Q_n^-(t).
 \label{eq:Q-}
}
The last terms in the r.h.s.\ of Eqs.\ (\ref{eq:Q+}) and (\ref{eq:Q-})
denote the remainder of particle pairs leaving the $n$-th shell during
$\tau_{n-1}$ and $\tau_{n+1}$, respectively.
Because $\Delta r_n\ll1$ and $\tau_n\ll1$,
(i) the Kramers-Moyal expansion
and (ii) addition and subtraction of Eqs.\ (\ref{eq:Q+}) and
(\ref{eq:Q-})
lead to
\EQA{
\frac{\PD}{\PD t}(Q_n^+ + Q_n^-)
&+& \frac{\PD}{\PD n}
\left(\frac{Q_n^+}{\tau_n}-\frac{Q_n^-}{\tau_n}\right)
= 0,
\label{eq:add-Q}
\\
\frac{\PD}{\PD t}(Q_n^+ - Q_n^-)
&+& \frac{\PD}{\PD n}
\left(\frac{Q_n^+}{\tau_n}+\frac{Q_n^-}{\tau_n}\right)
\nonumber
\\
&&
= -2\lambda^+\frac{Q_n^+}{T_c(\hat{r})}
+ 2\lambda^-\frac{Q_n^-}{T_c(\hat{r})}.
\label{eq:sub-Q}
}
In order to combine Eqs.\ (\ref{eq:add-Q}) and (\ref{eq:sub-Q}),
first,
divided by $\tau_n$ and differentiated with respect to $n$,
(\ref{eq:sub-Q}) leads to
\EQA{
\frac{\PD^2}{\PD n \PD t}
\left(\frac{Q_n^+}{\tau_n}-\frac{Q_n^-}{\tau_n}\right)
+ \frac{\PD}{\PD n}\left[
\frac{1}{\tau_n}\frac{\PD}{\PD n}
\left(\frac{Q_n^+}{\tau_n}+\frac{Q_n^-}{\tau_n}\right)
\right]
\nonumber
\\
=
\frac{\lambda}{T_c(\hat{r})}\frac{\PD Q_n}{\PD t}
-\frac{\delta}{T_c(\hat{r})}\frac{\PD}{\PD n}
\left(\frac{Q_n}{\tau_n}\right),
\label{eq:sub-Q2-rhs}
}
where $Q_n\equiv Q_n^+ + Q_n^-$,
$\lambda\equiv\lambda^+ + \lambda^-$,
and $\delta\equiv\lambda^+ - \lambda^-$.
Next, differentiating (\ref{eq:add-Q}) with respect to $t$ and substituting
 (\ref{eq:sub-Q2-rhs}) to it,
we obtain an equation for $Q_n$
around $\hat{r}$:
\EQA{
\frac{T_c(\hat{r})}{\lambda}
\frac{\PD^2 Q_n}{\PD t^2}
\EQAMK\EQAMK
+
\frac{\PD Q_n}{\PD t}
\nonumber
\\
=
\frac{T_c(\hat{r})}{\lambda}
\EQAMK\EQAMK
\frac{\PD}{\PD n}
\left[
\frac{1}{\tau_n}
\frac{\PD}{\PD n}
\left(\frac{Q_n}{\tau_n}\right)
\right]
+
\frac{\delta}{\lambda}\frac{\PD}{\PD n}
\left(\frac{Q_n}{\tau_n}\right).
\label{eq:T-Model-around-r}
}

Taking the limit as $\gamma\rightarrow 0$,
$r_n$ approaches $\hat{r}$.
And thus we obtain the following relations:
$Q_n(t)=\int_{r_n}^{r_{n+1}}P(r,t) dr\approx\gamma \hat{r}P(\hat{r},t)$,
$\PD/\PD n\approx\gamma \hat{r}\PD/\PD \hat{r}$,
and $\gamma/\tau_n\approx v(\hat{r})/\hat{r}$.
Replacing $\hat{r}$ with $r$,
we finally have
an equation for $P(r,t)$:
\EQA{
\frac{T_c(r)}{\lambda}
\frac{\PD^2 P}{\PD t^2}
+
\frac{\PD P}{\PD t}
\nonumber
&=&
\frac{T_c(r)}{\lambda}\frac{\PD}{\PD r}
\left(
v(r)
\frac{\PD}{\PD r}
\left[
v(r)P(r)
\right]
\right)
\\
&+&
\frac{\delta}{\lambda}\frac{\PD}{\PD r}
\left[
v(r)P(r)
\right].
\label{eq:T-Model-general}
}
Then Eq.\ (\ref{eq:T-Model-general}) with the scaling assumptions
yields
\EQA{
\frac{T_c(r)}{\lambda}
\EQAMK\EQAMK
\frac{\PD^2 P}{\PD t^2}
+
\frac{\PD P}{\PD t}
\nonumber
\\
=
\EQAMK\EQAMK
\frac{\PD}{\PD r}
\left[
D(r)r^{d-1}\frac{\PD}{\PD r}\left(\frac{P}{r^{d-1}}\right)
\right]
+\sigma\frac{\PD}{\PD r}
\left[v(r)P\right]
,
\label{eq:T-Model}
\qquad
}
where
$D(r)$ is Richardson's diffusion coefficient, $A\lambda^{-1}r^{2-g}$,
and
$\sigma=(d-2g+\delta)\lambda^{-1}\equiv\tilde{\sigma}\lambda^{-1}$.
The conservation of probability is satisfied at least for the similarity
solution.

In the limit of infinite speed and $\delta$-correlation,
the first term of Eq.\ (\ref{eq:T-Model}) disappears, which is the same
form as Palm's equation (see p.575 of \cite{MY1975}) 
\footnote{
Goto and Vassilicos derived the same equation \cite{GV2004}.
}.
Non-Richardson terms,
the first term in the l.h.s.\ and the last one in
the r.h.s.\ of Eq.\ (\ref{eq:T-Model}),
describe effects of
persistent separation.
The last term in the r.h.s.\ of Eq.\ (\ref{eq:T-Model})
is a drift term consistent with
the scaling assumptions.
The drift velocity is $-\sigma v(r)$ and, hence,
the direction is determined by $\tilde{\sigma}$
which consists of the ``scaling-determined'' part, $d-2g$,
and the ``dynamics-determined'' part, $\delta$.

Now, let us calculate the similarity solution of Eq.\ (\ref{eq:T-Model}).
To obtain it,
we introduce the  similarity variable $\xi$, 
\EQA{
\xi \equiv
\left(\frac{r}{\langle r^{2}\rangle^{1/2}}\right)^{g}
= C_R^{-g}\frac{\lambda}{At}r^g,
}
where we set  $\langle r^2\rangle^{1/2}=C_R(At/\lambda)^{1/g}$
\footnote{
$C_{R}^2(A/\lambda)^3/\varepsilon$ is the Richardson constant
in the Kolmogorov scaling, i.e., $g=2/3$.
}.
Substituting  $P(r,t)=F(\xi)\langle r^2\rangle^{-1/2}$ into
Eq.\ (\ref{eq:T-Model}),
we obtain an equation of
$F(\xi)=F(\eta/C_R^g)=\tilde{F}(\eta)$:
\EQA{
\frac{d^2 \tilde{F}}{d \eta^2}
+H_1(\eta) \frac{d \tilde{F}}{d \eta}
+H_2(\eta) \tilde{F}
=0,
\label{eq:F}
}
where
$\eta=C_R^{-g}\xi$,
\EQA{
H_1(\eta)
&=&
\frac{
2(1+g)\eta^2
- \lambda^2 g \eta
- \lambda^2 g^2 (2-\tilde{d})
}{
(\eta^2-\lambda^2 g^2) g \eta
},
\label{eq:def-H1}
\\
H_2(\eta)
&=&
\frac{
(1+g) \eta^2
- \lambda^2 g \eta
- \lambda^2 g^2 (1-\tilde{d})(1-g)
}{
(\eta^2-\lambda^2 g^2) g^2 \eta^2
},
\label{eq:def-H2}
\qquad
}
and $\tilde{d}=d-\tilde{\sigma}$.
Eq.\ (\ref{eq:F}) has three fixed singular points: 
$\xi=0$ and $\xi=\xi_{\pm}=\pm\lambda g/C_{R}^{g}$. 
Starting from an initial condition, where all particles occupy  a
small and compact region, the  solution is confined  in  $[0,\xi_+]$.
This finiteness of the similarity solution is due to the assumption of
a finite speed
and, thus, 
the maximum relative separation, $r_{\mathrm{max}}=\xi_{+}^g\langle
r^2\rangle^{1/2}=(Agt)^{1/g}$, exists 
\footnote{
 $r_{\mathrm{max}}$ is obtained by integrating $dr/dt=v(r)=Ar^{1-g}$ from $t=0$ to $t$.
}.

To see asymptotic behaviors of the similarity solution,
we consider two limits: the small ($\xi \ll1$) and
the
maximum
($\xi\sim\xi_{+}$) separation regimes.
In the small separation regime, $\xi\ll1$ or $r\ll \langle r^2\rangle^{1/2}$,
we obtain the following solution:  
\EQA{
F(\xi)\propto (\beta_{1}\xi)^{\beta_{2}}\exp(\beta_{1}\xi),
\label{eq:similarity-solution1}
}
where
$\beta_{1}=(d-\tilde{\sigma}-1)/g$ and $\beta_{2}=C_{R}^{g}/g^{2}$.
This solution is also a similarity solution of Palm's equation,
because the first term in Eq.\ (\ref{eq:T-Model}) can be  neglected 
 under the condition $r\ll \langle r^2\rangle^{1/2}$.
 We call this limit  the diffusive regime.

On the other hand, in the maximum separation regime,
the frontal edge of the PDF is abrupt at $r_{\mathrm{max}}$.
The functional form of the edge is approximated to the first order by
\EQA{
F(\xi)\propto(\xi_+ - \xi)^{\frac{\lambda-d}{2g}}.
\label{eq:similarity-solution2}
}
We call this limit  the telegraphic regime.
From this expression, it is clear that if $\lambda<d$,
$P(r,t)\rightarrow\infty$ as $r\rightarrow r_{\mathrm{max}}$.
That is, most of particle pairs are accumulated at the frontal edge,
where relative separations expand away without changing their directions.
However, this situation is unrealistic in real turbulence,
so that $\lambda$ is considered to be greater than $d$.

Control parameters of our model are $\lambda$ and
$\tilde{\sigma}$.
The functional form of the PDF,
$F(\xi)$,
is determined by 
these parameters: $\tilde{\sigma}$
controls mainly the diffusive regime,
and $\lambda$ does the telegraphic regime.
As $\lambda=1/\Ps^+ + 1/\Ps^-$, it represents the strength of 
persistency of moving direction.
On the other hand,
as $\tilde{\sigma}\lambda^{-1}$ is the coefficient of the drift term,
$\tilde{\sigma}$ represents total effects of persistent separations and
probabilistic transitions.
Because
persistent motions model the advection by coherent and self-similar
flows,
$\tilde{\sigma}$
seems to characterize the average effects of flow structures
such as coherent structures on dispersion processes.
In order to calculate the value of $\tilde{\sigma}$, we have to estimate
``dynamics-determined'' part of it, $\delta$.

To estimate $\delta$ from direct numerical simulations (DNS),
we use the PDF of exit-time \cite{ABCCV1997,BS2002a}.
Exit-time for particle-pairs experienced 
many turns  form an exponential tail
(for details see \cite{BS2002a}). 
In our case, the slope is evaluated with
Eq.\ (\ref{eq:T-Model}) in the limit of infinite
time, i.e., Palm's equation,
where the slope is related to $\delta$
\footnote{
In our case,
the asymptotic form of the PDF of exit-time $T_E$
from $r$ to $\rho r$ is
$\exp[
-\frac{1}{4}\frac{g}{2g-\delta}
j^2_{1-\delta/g,1}(1-\rho^{-g})
T_E \langle T_E(r;\rho)\rangle^{-1}
]$,
where $j_{\nu,n}$ is the $n$-th zero of the $\nu$-th order Bessel
function and $\langle T_E(r;\rho)\rangle$ is the mean exit-time from $r$
to $\rho r$.
}.
We can also estimate  $\delta$ directly from the separation PDF  
around $r\ll \langle r^2 \rangle ^{1/2}$.
In the 2D-IC case, 
Goto and Vassilicos estimated  $\alpha=2/(g-\delta)$ by fitting 
the similarity solution of  Palm's equation to the PDF \cite{GV2004}.
\begin{table}
\caption{\label{tab:delta}
 Estimated values of $\delta$ and $\tilde{\sigma}$ in 2D-IC and -FC turbulences.
 }
\begin{ruledtabular}
\begin{tabular}{cccc}
Type & Method & $\delta$ & $\tilde{\sigma}$ \\
\hline
2D-IC\footnote{From DNS results by Boffetta and Sokolov
 \cite{BS2002a}. $\mathrm{Slope}\approx-0.3$.} &
 Exit-time PDF & -1.48 & -0.81 \\
2D-IC\footnote{From DNS results by Goto and Vassilicos
 \cite{GV2004}. $\alpha\approx1.3$.} & Separation PDF & -0.87
 & -0.20 \\
2D-FC\footnote{From our DNS with resolution $2048^2$ \cite{OT2005a}.} & Exit-time PDF &
 -0.71 & 0.49 \\
\end{tabular}
\end{ruledtabular}
\end{table}

Table \ref{tab:delta} shows the estimated values of $\delta$ and $\tilde{\sigma}$
for 2D-IC and -FC turbulences.
These results indicate that 
 the drift term of (\ref{eq:T-Model}) 
enhances diffusion in the 2D-IC case
but suppresses diffusion in the 2D-FC case;  
Compression  of relative separations in 2D-IC turbulence
but expansion of them  in 2D-FC turbulence
are comparatively restricted, respectively.
 This remarkable feature is considered to be induced by the 
difference in flow structures between 2D-IC and -FC turbulences:
 ``cat's eye in a cat's eye'' structures \cite{GV2004} and  
string-like structures \cite{TM2003}.
We, therefore, expect that 
$\tilde{\sigma}$ can characterize coherent 
structures.

\begin{figure}[t]
 \includegraphics[width=8.5cm]{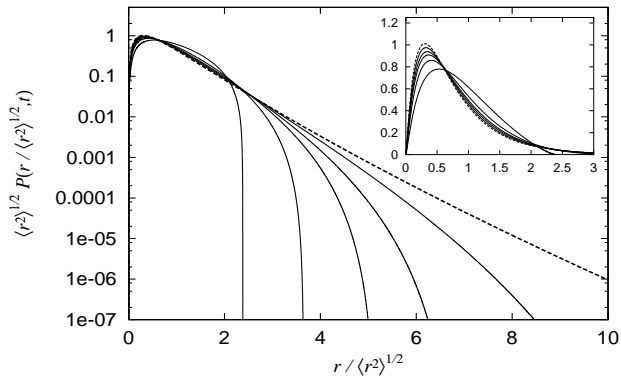}
 \caption{\label{fig:TM-fPDF}
 PDFs of relative separation
 for the 2D-IC case.
 Solid lines represent similarity solutions of our model with
 $\lambda=4$, $6$, $8$, $10$, and $15$ from the curve with the smallest cut-off scale to
 that of the largest.
 The dashed line represents
 $\lambda=\infty$, or 
 the similarity solution of Palm's equation (\ref{eq:similarity-solution1}).
 $\tilde{\sigma}$ is fixed for $1/3$: $d=2$, $g=2/3$, and $\delta=-1$.
 The inset is the same plot in linear scale.
 Lines correspond to
 $\lambda=4$, $6$, $8$, $10$, $15$, and $\infty$ from the curve with the
 lowest peak value
 to that of the highest.
 }
\end{figure}
In Fig.\ \ref{fig:TM-fPDF},  the similarity solution $F(\xi)$ obtained numerically
 for various values of  $\lambda$ is shown.
 A cut-off scale corresponding to $r_{\mathrm{max}}$ can be seen.
 Our similarity  solution  approaches Palm's one as
 $\lambda$ gets larger.
 However, even for large $\lambda$,
 the difference between them in the tail part
 is so evident that 
 effects of persistent separation
are not negligible.

In summary, 
we derived a telegraph equation with scale-dependent coefficients,
into which finite separation speed and self-similarity are
incorporated,
by employing multiple-scale consideration in scale-space.
Then we obtained a similarity solution of it.
In the 
diffusive regime,
$\xi \ll 1$, the similarity solution 
coincides with  that of Richardson's diffusion
equation with the drift term, i.e., Palm's equation; 
in the telegraphic regime,
$\xi \sim \xi_\mathrm{max}$, the finiteness of separation speed is realized
and the separation PDF is abrupt at $r_{\mathrm{max}}$.
Therefore, finite separation speed is crucial for description of the tail part
of the separation PDF unless relative velocity is $\delta$-correlated in
time.

The drift term of Eq.\ (\ref{eq:T-Model})
 is induced by 
the deviation of the difference of persistency between expansion and
compression of relative separation, $\delta$,
from ``scaling-determined'' value, $2g-d$.
The direction of the drift is determined by the sign of
$-\tilde{\sigma}=(2g-d)-\delta$.
We estimated the value for two 2-D turbulences,
inverse cascade (IC) and free convection (FC),
and found that positive and negative drift is imposed in the 2D-IC and
-FC cases, respectively.
We conjecture that this remarkable difference corresponds to
the different types of coherent structures in the background flow.
We need more precise investigations to obtain an evidence for this.

We neglected two significant effects:  
distribution of separation speed
  and intermittency.
However, intermittency of relative velocity is
negligible in the two 2D turbulences dealt with in this letter. 
Besides, not the distribution but the finiteness and self-similarity of 
separation speed is crucial for the existence of the cut-off of
the separation PDF. We are expecting experiments which can resolve 
the tail part of the separation PDF. 

 \begin{acknowledgments}
  This work is supported by the Grant-in-Aid for the 21st Century COE
  ``Center for Diversity and Universality in Physics'' from the Ministry
  of Education, Culture, Sports, Science and Technology (MEXT) of Japan.
  Numerical computation in this work was carried out on a NEC SX-5 at the
  Yukawa Institute Computer Facility.
 \end{acknowledgments}


\begin{thebibliography}{18}
\expandafter\ifx\csname natexlab\endcsname\relax\def\natexlab#1{#1}\fi
\expandafter\ifx\csname bibnamefont\endcsname\relax
  \def\bibnamefont#1{#1}\fi
\expandafter\ifx\csname bibfnamefont\endcsname\relax
  \def\bibfnamefont#1{#1}\fi
\expandafter\ifx\csname citenamefont\endcsname\relax
  \def\citenamefont#1{#1}\fi
\expandafter\ifx\csname url\endcsname\relax
  \def\url#1{\texttt{#1}}\fi
\expandafter\ifx\csname urlprefix\endcsname\relax\def\urlprefix{URL }\fi
\providecommand{\bibinfo}[2]{#2}
\providecommand{\eprint}[2][]{\url{#2}}

\bibitem[{\citenamefont{Low and Klessen}(2004)}]{MK2004}
\bibinfo{author}{\bibfnamefont{M.-M.} \bibnamefont{Mac~Low}} \bibnamefont{and}
  \bibinfo{author}{\bibfnamefont{R.~S.} \bibnamefont{Klessen}},
  \bibinfo{journal}{Rev. Mod. Phys.} \textbf{\bibinfo{volume}{76}},
  \bibinfo{pages}{125} (\bibinfo{year}{2004}).

\bibitem[{\citenamefont{Richardson}(1926)}]{R1926}
\bibinfo{author}{\bibfnamefont{L.~F.} \bibnamefont{Richardson}},
  \bibinfo{journal}{Proc. R. Soc. A} \textbf{\bibinfo{volume}{110}},
  \bibinfo{pages}{709} (\bibinfo{year}{1926}).

\bibitem[{\citenamefont{Monin and Yaglom}(1975)}]{MY1975}
\bibinfo{author}{\bibfnamefont{A.~S.} \bibnamefont{Monin}} \bibnamefont{and}
  \bibinfo{author}{\bibfnamefont{A.~M.} \bibnamefont{Yaglom}},
  \emph{\bibinfo{title}{Statistical Fluid Mechanics}}, vol.~\bibinfo{volume}{2}
  (\bibinfo{publisher}{MIT Press}, \bibinfo{year}{1975}).

\bibitem[{\citenamefont{Sawford}(2001)}]{S2001}
\bibinfo{author}{\bibfnamefont{B.}~\bibnamefont{Sawford}},
  \bibinfo{journal}{Annu. Rev. Fluid Mech.} \textbf{\bibinfo{volume}{33}},
  \bibinfo{pages}{289} (\bibinfo{year}{2001}).

\bibitem[{\citenamefont{Jullien et~al.}(1999)\citenamefont{Jullien, Paret, and
  Tabeling}}]{EXPS}
\bibinfo{author}{\bibfnamefont{M.~C.} \bibnamefont{Jullien}},
  \bibinfo{author}{\bibfnamefont{J.}~\bibnamefont{Paret}}, \bibnamefont{and}
  \bibinfo{author}{\bibfnamefont{P.}~\bibnamefont{Tabeling}},
  \bibinfo{journal}{Phys. Rev. Lett.} \textbf{\bibinfo{volume}{82}},
  \bibinfo{pages}{2872} (\bibinfo{year}{1999});
%
\bibinfo{author}{\bibfnamefont{S.}~\bibnamefont{Ott}} \bibnamefont{and}
  \bibinfo{author}{\bibfnamefont{J.}~\bibnamefont{Mann}}, \bibinfo{journal}{J.
  Fluid Mech.} \textbf{\bibinfo{volume}{422}}, \bibinfo{pages}{207}
  (\bibinfo{year}{2000});
%
\bibinfo{author}{\bibfnamefont{G.}~\bibnamefont{Boffetta}} \bibnamefont{and}
  \bibinfo{author}{\bibfnamefont{I.~M.} \bibnamefont{Sokolov}},
  \bibinfo{journal}{Phys. Rev. Lett.} \textbf{\bibinfo{volume}{88}},
  \bibinfo{pages}{094501} (\bibinfo{year}{2002}{\natexlab{b}});
%
\bibinfo{author}{\bibfnamefont{P.~K.} \bibnamefont{Yeung}} \bibnamefont{and}
  \bibinfo{author}{\bibfnamefont{M.~S.} \bibnamefont{Borgas}},
  \bibinfo{journal}{J. Fluid Mech.} \textbf{\bibinfo{volume}{503}},
  \bibinfo{pages}{93} (\bibinfo{year}{2004}).

\bibitem[{\citenamefont{Boffetta and Sokolov}(2002{\natexlab{a}})}]{BS2002a}
\bibinfo{author}{\bibfnamefont{G.}~\bibnamefont{Boffetta}} \bibnamefont{and}
  \bibinfo{author}{\bibfnamefont{I.~M.} \bibnamefont{Sokolov}},
  \bibinfo{journal}{Phys. Fluids} \textbf{\bibinfo{volume}{14}},
  \bibinfo{pages}{3224} (\bibinfo{year}{2002}{\natexlab{a}}).

\bibitem[{\citenamefont{Goto and Vassilicos}(2004)}]{GV2004}
\bibinfo{author}{\bibfnamefont{S.}~\bibnamefont{Goto}} \bibnamefont{and}
  \bibinfo{author}{\bibfnamefont{J.~C.} \bibnamefont{Vassilicos}},
  \bibinfo{journal}{New J. Phys.} \textbf{\bibinfo{volume}{6}},
  \bibinfo{pages}{65} (\bibinfo{year}{2004}).

\bibitem[{\citenamefont{Sokolov}(1999)}]{S1999}
\bibinfo{author}{\bibfnamefont{I.~M.} \bibnamefont{Sokolov}},
  \bibinfo{journal}{Phys. Rev. E} \textbf{\bibinfo{volume}{60}},
  \bibinfo{pages}{5528} (\bibinfo{year}{1999}).

\bibitem[{\citenamefont{Toh and Matsumoto}(2003)}]{TM2003}
\bibinfo{author}{\bibfnamefont{S.}~\bibnamefont{Toh}} \bibnamefont{and}
  \bibinfo{author}{\bibfnamefont{T.}~\bibnamefont{Matsumoto}},
  \bibinfo{journal}{Phys. Fluids} \textbf{\bibinfo{volume}{15}},
  \bibinfo{pages}{3385} (\bibinfo{year}{2003}).

\bibitem[{\citenamefont{D\'{a}vila and Vassilicos}(2003)}]{DV2003}
\bibinfo{author}{\bibfnamefont{J.}~\bibnamefont{D\'{a}vila}} \bibnamefont{and}
  \bibinfo{author}{\bibfnamefont{J.~C.} \bibnamefont{Vassilicos}},
  \bibinfo{journal}{Phys. Rev. Lett.} \textbf{\bibinfo{volume}{91}},
  \bibinfo{pages}{144501} (\bibinfo{year}{2003}).

\bibitem[{\citenamefont{Sokolov et~al.}(2000)\citenamefont{Sokolov, Klafter,
  and Blumen}}]{SKB2000}
\bibinfo{author}{\bibfnamefont{I.~M.} \bibnamefont{Sokolov}},
  \bibinfo{author}{\bibfnamefont{J.}~\bibnamefont{Klafter}}, \bibnamefont{and}
  \bibinfo{author}{\bibfnamefont{A.}~\bibnamefont{Blumen}},
  \bibinfo{journal}{Phys. Rev. E} \textbf{\bibinfo{volume}{61}},
  \bibinfo{pages}{2717} (\bibinfo{year}{2000}).

\bibitem[{\citenamefont{Goldstein}(1951)}]{G1951}
\bibinfo{author}{\bibfnamefont{S.}~\bibnamefont{Goldstein}},
  \bibinfo{journal}{Quart. J. Mech. Appl. Math.} \textbf{\bibinfo{volume}{4}},
  \bibinfo{pages}{129} (\bibinfo{year}{1951}).

\bibitem[{\citenamefont{Bakunin}(2003)}]{B2003}
\bibinfo{author}{\bibfnamefont{O.~G.} \bibnamefont{Bakunin}},
  \bibinfo{journal}{Phys. Uspekhi} \textbf{\bibinfo{volume}{46}},
  \bibinfo{pages}{309} (\bibinfo{year}{2003}).

\bibitem[{\citenamefont{Holmes}(1993)}]{H1993}
\bibinfo{author}{\bibfnamefont{E.~E.} \bibnamefont{Holmes}},
  \bibinfo{journal}{Am. Nat.} \textbf{\bibinfo{volume}{142}},
  \bibinfo{pages}{779} (\bibinfo{year}{1993}).

\bibitem[{\citenamefont{Artale et~al.}(1997)\citenamefont{Artale, Boffetta,
  Celani, Cencini, and Vulpiani}}]{ABCCV1997}
\bibinfo{author}{\bibfnamefont{V.}~\bibnamefont{Artale}},
  \bibinfo{author}{\bibfnamefont{G.}~\bibnamefont{Boffetta}},
  \bibinfo{author}{\bibfnamefont{A.}~\bibnamefont{Celani}},
  \bibinfo{author}{\bibfnamefont{M.}~\bibnamefont{Cencini}}, \bibnamefont{and}
  \bibinfo{author}{\bibfnamefont{A.}~\bibnamefont{Vulpiani}},
  \bibinfo{journal}{Phys. Fluids} \textbf{\bibinfo{volume}{9}},
  \bibinfo{pages}{3162} (\bibinfo{year}{1997}).

\bibitem[{\citenamefont{Ogasawara and Toh}(2005)}]{OT2005a}
\bibinfo{author}{\bibfnamefont{T.}~\bibnamefont{Ogasawara}} \bibnamefont{and}
  \bibinfo{author}{\bibfnamefont{S.}~\bibnamefont{Toh}}, in
  \emph{\bibinfo{booktitle}{IUTAM Symposium on Elementary Vortices and Coherent
  Structures: Significance in Turbulence Dynamics}}, edited by
  \bibinfo{editor}{\bibfnamefont{S.}~\bibnamefont{Kida}}
  (\bibinfo{publisher}{Springer}, \bibinfo{year}{2005}), p.
  \bibinfo{pages}{143}, \bibinfo{note}{in press}.

\bibitem[{\citenamefont{Toh and Suzuki}(1994)}]{TS1994}
\bibinfo{author}{\bibfnamefont{S.}~\bibnamefont{Toh}} \bibnamefont{and}
  \bibinfo{author}{\bibfnamefont{E.}~\bibnamefont{Suzuki}},
  \bibinfo{journal}{Phys. Rev. Lett.} \textbf{\bibinfo{volume}{73}},
  \bibinfo{pages}{1501} (\bibinfo{year}{1994}).

\end{thebibliography}

\end{document}